# Temperature-dependent Multi-well Free-energy Landscape for Phase Transitions: PbTiO$_3$ as a Prototype

Yi Wang, Tiannan Yang, Shun-Li Shang, Long-Qing Chen*, and Zi-Kui Liu

*Materials Science and Engineering, The Pennsylvania State University, University Park, PA 16802, USA.*

It has been a long challenge to analytically construct the quantitative temperature-dependent multi-well free-energy landscape over the space of order parameters describing phase transitions and associated critical phenomena. Here we propose a simple analytical model for the free energy landscape based on *a priori* concept of Boltzmann thermal mixing among multiple parabolic potentials representing different energetically degenerated ground states in contrast to the popular Landau theory using the high-temperature disordered state as a reference. The model recovers both the Weiss molecular field theory and the temperature dependent behaviors of the second order Landau coefficient. It is rather remarkable that such a simple analytical expression can describe a wide variety of properties across the ferroelectric phase transition in PbTiO$_3$, including the temperature dependences of the spontaneous polarization, the heat capacities, the permittivity tensor, and the lattice parameters. The approach also allows parametrization of the free energy function across phase transitions based directly on the 0 K thermodynamics data that can be obtained from DFT calculations. We anticipate that the approach can be equally applied to other types of critical phenomena, e.g., the superconducting phase transition, the metal-insulator transition, and the magnetic transitions.

*lqc3@psu.edu

# 1    Introduction

All phase transitions, such as ferroelectric, magnetic, and superconductive phase transitions, are characterized by the abrupt or dramatic changes in many of its physical properties at the transition temperature. The thermodynamics of a phase transition is often described by a multi-well free-energy landscape of order parameters characterizing the differences between the phases across the transition [1–7], e.g., using either the Landau expansion [8,9] or the piecewise polynomials [10]. Many phase transitions exhibit similar behaviors [11,12] independent of the dynamical details of a material, i.e., universality as defined by Kadanoff [11].

The purpose of the present study is to develop an alternative phenomenological theory to the widely adopted Landau theory [8,9] for analyzing the phase transitions such as magnetic transition, the superconducting transition, metal-insulator transition, and the ferroelectric transition, including the temperature dependences of the corresponding order parameters and thermal properties. In contrast to the Landau theory which builds its analytical potential energy function using the high temperature disordered state as reference, we employ the single domain, ordered, and equilibrium ground state as the reference state for each temperature. One of the main advantages of our approach is the fact all the thermodynamic properties of the equilibrium ordered state as a function of temperature can in principle be computed using a combination of density functional theory and phonon calculations [13] whereas it is challenging if not impossible to compute thermodynamic properties of the high-temperature disordered reference state directly from density function theory. To obtain an analytical free-energy function with multiple wells representing multiple energetically degenerate ordered states as a function of order parameters and temperature, we employ the Boltzmann thermal mixing ([14–17]) of multiple parabolic wells with different well centers representing the ground states at each temperature. Such analytical free energy functions are essential for building thermodynamic and mesoscale material models like phase-field models [18] within a multiscale theoretical framework.

The present study is developed in the framework of mean-field theory [19,20] (also named as molecular field theory) by which the effect of all the other individuals on any given individual is approximated by an averaged field. Based on the fact that the ferroelectric-paraelectric phase transition obeys that Curie-Weiss law [21], it is natural to treat the ferroelectric-paraelectric phase transition as the disordering of the local polarizations, resembling the case of ferromagnetic-paramagnetic phase transion [22,23] or the ferroelectric-paraelectric phase transition in relaxor [24,25].

We postulate that a macroscopic ferroelectirc system is made of a massive number of small polar atomic clusters (PAC's) that have fixed size and are quantized by the their electric dipole directions. We further assume the sizes of the PAC's are constant and leave it as a phenological fitting parameter when needed. The ground state of the system is a condensate state that the dipole moments of all PAC's are frozen into the same directions, resulting in a single domain, uniform, and equilibrium ordered state. Being the same as the ferromagnetic-paramagnetic phase transion [22,23] or the relaxor, the local dipole moments of the PAC's are not zero in the paraelectric phase and the phase transition is due to the microscopic disordering of the dipole directions of the PAC's.

To formulate the energetics of the ferroelectric-paraelectric phase transition, it is more convenient to use the polarization which is the quotient of the dipole moment divived by the volume of the PAC. Let us use the symbol $\Xi_\sigma$ to represent the polarization of a local PAC where the index σ labels the direction of the polarization. According to the mean-field theory, the averaged field seen by a local PAC is proportional to the macroscopic polarization $\Xi$ which is the thermal average of $\Xi_\sigma$. The idea can be implemented by making a virtual perturbation to the ground state of the system, designated by $\Xi_\sigma$, to the macroscopic state, designated by $\Xi$. By a Taylor expansion to the second order, we propose to write the energy of a local PAC in the parabolic form as

$$\mathcal{H}^\sigma(\Xi, T) = \mathcal{H}(\Xi_\sigma, T) + \Delta\mathcal{H}(\Xi - \Xi_\sigma) \quad \text{Eq. 1}$$

where $\mathcal{H}(\Xi_\sigma, T)$ is the free energy of the homogeneous ground state at equilibrium, $T$ is the temperature, and

$$\Delta\mathcal{H}(\Xi - \Xi_\sigma) = \frac{1}{2}(\Xi - \Xi_\sigma) \cdot \left.\frac{\delta^2 \mathcal{H}}{\delta \Xi^2}\right|_{\Xi=\Xi_\sigma} \cdot (\Xi - \Xi_\sigma)$$

$$= \frac{1}{2}\Xi \cdot \Lambda_\sigma \cdot \Xi - \Xi \cdot \Lambda_\sigma \cdot \Xi_\sigma + \frac{1}{2}\Xi_\sigma \cdot \Lambda_\sigma \cdot \Xi_\sigma \quad \text{Eq. 2}$$

where $\Lambda_\sigma = \left.\frac{\delta^2 \mathcal{H}}{\delta \Xi^2}\right|_{\Xi=\Xi_\sigma}$ represents the second order derivative tensor of the free energy of the ground state at $\Xi = \Xi_\sigma$ and the dot symbol " $\cdot$ " represents the algebraic operation of dot product. From Eq. 2, one can see the quadratic term ($\frac{1}{2}\Xi \cdot \Lambda_\sigma \cdot \Xi$) and the long range field term ($\Xi \cdot \Lambda_\sigma$) based on them one can recover the major results of the Bragg-Williams approximation [26]. Furthermore, the term $\frac{1}{2}\Xi_\sigma \cdot \Lambda_\sigma \cdot \Xi_\sigma$ makes the reference energy to zero which is more convenient for adding other contributions such as those of lattice vibration and 0 K energetics.

We ascribe the ferroelectric-paraelectric phase transition to the microscopic polarization disordering of the PAC's. As a proof of concept, we further assume that all $\mathcal{H}(\Xi_\sigma, T)'s$ introduced in Eq. 1 are independent of the quantized index σ, i.e., $\mathcal{H}(\Xi_\sigma, T)'s = \mathcal{H}_0(T)$ which is generally true when all the ground states are degenerate (i.e., possessing the same structure except for different orientations). At finite temperature, the partition function for the macroscopic system is a summation over multiple parabola with their centers located at $\Xi_\sigma$

$$Z = \sum_{\sigma} exp[-\beta \mathcal{H}^{\sigma}(\Xi, T)] \quad \text{Eq. 3}$$

where $\beta = 1/(k_B T)$ ($k_B$ being the Boltzmann constant). The Gibbs energy can then be derived using $G = -\frac{lnZ}{\beta} = \mathcal{H}_0(T) + \Delta G(\Xi, T)$ where

$$\Delta G(\Xi, T) = -\frac{1}{\beta} ln\left\{\sum_{\sigma} exp[-\beta \Delta \mathcal{H}(\Xi - \Xi_{\sigma})]\right\} \quad \text{Eq. 4}$$

$\Delta G(\Xi, T)$ accounts for the extra Gibbs energy with respect to the referenced ground state. Minimizing the extra Gibbs energy, one can find that the optimized $\Xi$ equals to the thermal average of the individual $\Xi_{\sigma}$.

Eq. 4 can be simplified when the second order derivatives of the Gibbs energy is a scalar, i.e., $\frac{\delta^2 \mathcal{H}}{\delta \Xi^2}\Big|_{\Xi=\Xi_{\sigma}} = \chi \mathbf{I}$ where $\chi$ is a scaler number and $\mathbf{I}$ is a unit matrix. As a matter of fact, when $T \to 0$, the extra Gibbs energy $\Delta G(\Xi, T)$ in Eq. 4 has multiple (2D) equivalent minima located at $\Xi_{\sigma} = \pm X_0 \hat{x}_g$ ($g$ = 1, 2, ..., $D$), where $\hat{x}_g$ is a unit vector. We further limit our consideration to the case of $\hat{x}_g \cdot \hat{x}_i = \delta_{gi}$ with $g, i$ = 1, 2, ..., $D$ while $D$ is apparently the dimensionality of the studied problem.

Depending on the problem, we are often interested in the evolution of $\Delta G(\Xi, T)$ within a cross section with dimensionality of $D_s$, then Eq. 4 can be simplified using $\hat{x}_i \cdot \hat{x}_g = \delta_{ig}$, resulting in

$$\Delta G(\xi, T) = D k_B T_k \left\{ \frac{1}{2} \sum_{i=1}^{D_s} \xi_i^2 + \frac{1}{2} - \frac{t}{D} ln\left[ 2(D - D_s) + \sum_{i=1}^{D_s} 2cosh(\xi_i \frac{D}{t}) \right] \right\} \quad \text{Eq. 5}$$

where $\xi_i = \Xi \cdot \hat{x}_i / X_0$ is the reduced order parameter which would reach ±1 at 0 K, $t = T/T_k$ is the reduced temperature, and $T_k = \chi X_0^2 / (D k_B)$ is a characteristic temperature which is a reminiscent of the Curie temperature [27,28] in the Ginzburg-Landau expansion [8,9]. It can be shown that $\frac{\partial^2 \Delta G}{\partial \xi_i^2} = D k_B \frac{T_k}{T}(T - T_k)$ at $\xi_i = 0$.

It is noted that at $D = 1$ and $D_s = 1$, the present approach is able to recover the result of magnetization from the Weiss molecular field theory [20] and the major results of free-energy from the Bragg-Williams approximation [26].

In this work, we will use PbTiO$_3$ as the prototype by treating $D=D_s=3$. In this special case, Eq. 5 can be explicitly written as

$$\Delta G(\xi, T) = 3k_B T_k \left\{ \frac{1}{2}\sum_{i=1}^{3} \xi_i^2 + \frac{1}{2} - \frac{t}{3} \ln\left[\sum_{i=1}^{3} 2\cosh(\xi_i \frac{3}{t})\right] \right\} \qquad \text{Eq. 6}$$

Figure 1a examines the general dependence of the extra Gibbs energy on the temperature and order parameter assuming the system is initially orientated along the $z$ direction (noted as $\xi = \xi_3$). To exhibit different physics from different approaches, the extra Gibbs energies from the Landau expansion are also plotted in Figure 1b for comparison based on the parameters of Haun et al. [28]. The extra Gibbs energy modelled by the present approach shows a monotonic decrease with increasing temperature, i.e., a positive extra entropy which is a common sense of thermodynamics. In comparison, the Landau expansion shows a monotonic increase with increasing temperature due to its usage of the paraelectric state as the reference state, while the present approach uses the ground state as the reference state. The implication of this change of reference state is profound as the Gibbs energy of the ground state can be predicted from first-principles-based phonon calculations, while the Gibbs energy of the paraelectric state cannot be accurately predicted at present due to the existence of imaginary phonon modes for most of the high temperature phases.

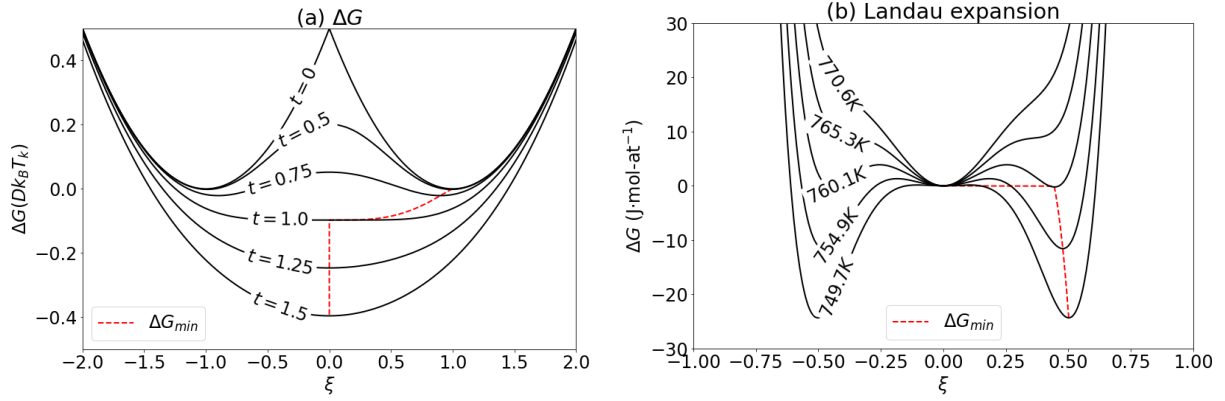

Figure 1. The temperature dependences of extra Gibbs energy. (a) the extra Gibbs energy of present approach; (b) the extra Gibbs energy from the Landau expansion using parameters from Haun et al. [28] for PbTiO$_3$. The dashed lines mark the temperature dependences of the optimized order parameter and global minima of the extra Gibbs energy.

The order parameter $\xi$ plays the role of the effective *long range field* which effectively imposes the energy penalty when a PAC switches its polar doirection. The optimized $\xi_j$ along a given axis *j* can be obtained by solving $\partial \Delta G(\xi, T)/\partial \xi_j = 0$ and one gets

$$\xi_j = \frac{sinh(\xi_j \frac{3}{t})}{\sum_{i=1}^{3} cosh(\xi_i \frac{3}{t})} \quad (j = 1,2,3) \qquad \text{Eq. 7}$$

Figure 2a illustrates the temperature evolution (in terms of the reduced temperature *t*) of the optimized order parameter along z direction (noted as $\xi = \xi_3$) assuming the system is initially orientated along the *z* direction. A salient feature is that the value of the optimized order parameter exponentially vanishes when approaching to the critical temperature.

With the optimized order parameter in hand, we can calculate the temperature dependence of spontaneous polarization. With *D*=3, the initial polarization $\Xi_\sigma$ introduced in Eq. 4 can point to six direction, i.e., the summation over $\sigma$ covers $(\pm 1,0,0)$, $(0,\pm 1,0)$, and $(0,0,\pm 1)$. Combining Eq. 4 and Eq. 6, one can find the phase fraction along each direction as

$$f_\sigma = \frac{exp(\xi_\sigma \frac{3}{t})}{\sum_{i=1}^{3} cosh(\xi_i \frac{3}{t})} \qquad \text{Eq. 8}$$

where $\sigma = \pm 1, \pm 2,$ and $\pm 3$ are used to note the 6 degenerate ground states with polarization along $\pm \hat{x}, \pm \hat{y},$ and $\pm \hat{z}$ directions, respectively and we have used the notation $\xi_{-j} = -\xi_j$ for brevity. As a result, the spontaneous polarization along a given axis $j$ can be calculated by

$$P_j = f_j P_0(T) + f_{-j}[-P_0(T)] = P_0(T)\xi_j \qquad \text{Eq. 9}$$

where $P_0$ represents the polarization magnitude of the ground state. Here we want to point out that $P_0$ is an intrinsic property that is a reminiscent of the concept of local spin of the Ising model for ferromagnetic-paramagnetic transition [22].

As a result, the ferroelectric-paraelectric transition in of PbTiO$_3$ should be understood as the disordering in terms of the local polarization or crystal orientation, as previously demonstrated by the *ab initio* molecular dynamics simulations of Fang et al. [29]. This understanding is well supported by the x-ray absorption fine-structure (XAFS) measurements of [30]. The measurements of Sicron et al. [30] showed that (i) the Pb/Ti atoms were displaced relative to the oxygen octahedra both below and above the transition temperature and ii) the local unit cells remained tetragonal over the entire temperature range of 0-1000 K.

As about the temperature dependence of $P_0$ for PbTiO$_3$, we assume is proportional to the Ti atom displacement measured by Sicron et al. [30]. Using $T_k$=764, the spontaneous polarization of PbTiO$_3$ has been calculated and Figure 2b compared the theoretical spontaneous polarization with the data reported by Haun er al. [28].

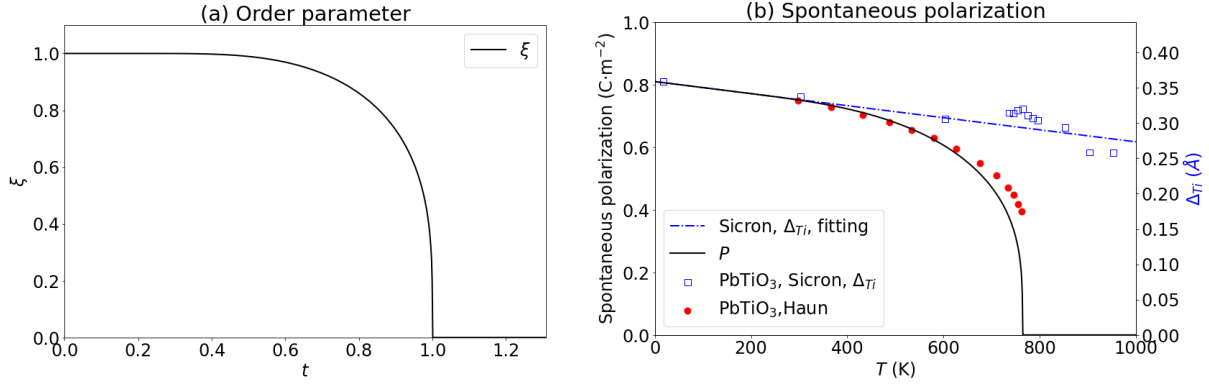

Figure 2. (a) The temperature evolution of the optimized order parameter. (b) Spontaneous polarization for PbTiO$_3$ where the theoretical value is represented by the solid line, the reported data reported by Haun er al. [28] are represented by solid circles. The dashed line represents the a linear fitting for the temperature dependence of the Ti displacement ($\Delta_{Ti}$, open squares) along the $c$ direction by the XAFS measurements of Sicron et al. [30] that are used to calibrate the temperature dependence of the polarization ($P_j^0$ as introduced in Eq. 9) of the ground state.

Along the same route, we can also calculate the temperature dependence of the $c$ and $a$ lattice parameters (assuming the initial polarization is along the $c$ direction) across the phase transition of PbTiO$_3$ by

$$[a \quad a \quad c] = (f_1 + f_{-1} \quad f_2 + f_{-2} \quad f_3 + f_{-3}) \cdot \begin{bmatrix} c^0(T) & a^0(T) & a^0(T) \\ a^0(T) & c^0(T) & a^0(T) \\ a^0(T) & a^0(T) & c^0(T) \end{bmatrix} \quad \text{Eq. 10}$$

where $c^0$ and $a^0$ are the lattice parameters of the ground state along the $c$ and $a$ directions, respectively. In this work, $c^0$ and $a^0$ have been determined by a linear fitting of the measured local lattice parameters by the XAFS measurements of Sicron et al. [30]. Figure 3 compares the calculated $c$ and $a$ lattice parameters with the measured data reported by Haun et al. [28] and Mabud et al. [31].

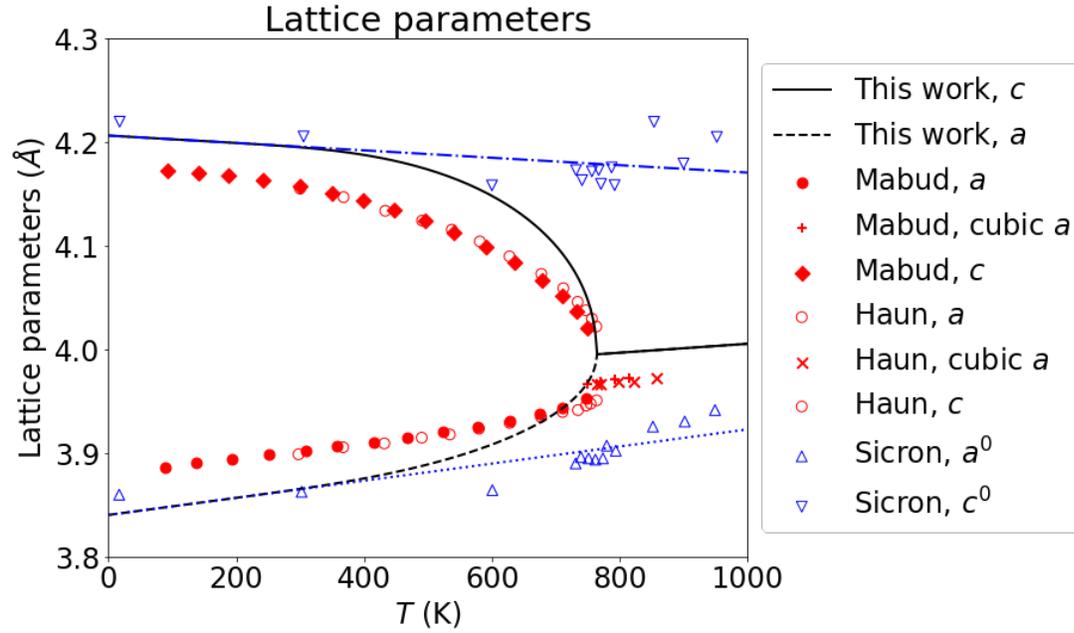

Figure 3. Temperature evolution of the lattice parameters of PbTiO$_3$. The solid and dashed lines are the calculated $c$ and $a$ lattice parameters, respectively; The dot-dashed and dotted lines represent the linear fittings for the measured temperature dependences of the local $c$ (down triangles) and $a$ (up triangles) lattice parameters of Sicron et al. [30], respectively. The reported $c$ and $a$ data by Haun er al. [28] are represented by the signs of solid circles and crosses. The measured $c$ and $a$ data reported by Mabud et al. [31] are represented by the signs of open circles and pluses.

The proposed approach can be also used to calculate the inverse relative electric permittivity. The formulation can be derived by finding the second order derivative, noted as $A_{ij}$ below, of the volume density of the extra Gibbs energy against the spontaneous polarization. The extra Gibbs energy has been given in Eq. 6 and the spontaneous polarization has been formulated in Eq. 9 and plotted in Figure 2b. One has

$$A_{ij} = \varepsilon_0 \left[\frac{\partial^2 (\Delta G/V_{PAC})}{\partial P_i \partial P_j}\right]_T = \frac{\varepsilon_0}{V_{PAC}} \sum_{k,l=1,2,3} \frac{\partial^2 \Delta G(\xi,T)}{\partial \xi_k \partial \xi_l} \frac{\partial \xi_k}{\partial P_i} \frac{\partial \xi_l}{\partial P_j} \quad \text{Eq. 11}$$

$$= \frac{3 k_B T_k \varepsilon_0}{V_{PAC}[P_0(T)]^2} \left\{ \delta_{ij} \left[1 - \frac{3}{t} \frac{\cosh\left(\xi_i \frac{3}{t}\right)}{\sum_{k=1}^3 \cosh(\xi_k \frac{3}{t})}\right] + \frac{3}{t} \frac{\sinh(\xi_i \frac{3}{t})\sinh(\xi_j \frac{3}{t})}{\left[\sum_{k=1}^3 \cosh(\xi_k \frac{3}{t})\right]^2} \right\}$$

where $\varepsilon_0$ is the vacuum permittivity. It is noted that $V_{PAC}$ in Eq. 11 is the volume of the PAC. $V_{PAC}$ can be estimated from the Curie constant $C$ [28] as follows. Consider around $T \to T_k$ in the paraelectric phase where $\xi_i = 0$. From Eq. 11, one gets

$$A_{33} = \frac{3 k_B T_k \varepsilon_0}{V_{PAC}[P_0(T)]^2} \left(\frac{T - T_k}{T}\right) = \frac{T - T_k}{C} \quad \text{Eq. 12}$$

which gives rise to $V_{PAC}$ as

$$V_{PAC} = \frac{3 k_B \varepsilon_0 C}{[P_0(T_k)]^2} \quad \text{Eq. 13}$$

Using $C=1.50\times10^5$ K as determined by Haun et al. [28], $P_0=0.6627$ C/m² estimated from Figure 2b, together with knowing $k_B=1.380649\times10^{-23}$ J/K and $\varepsilon_0=8.854\times10^{-12}$ F/m, one gets $V_{PAC} = 125.3\times10^{-30}$ m³. This is magically the volume of two perovskite unit cells of PbTiO$_3$ by Haun et al. [28] at the the transition temperature.

The relative electric permittivity can be obtained as matrix inverse of $A_{ij}$, i.e.

$$\varepsilon_{ij} = [\mathbf{A}^{-1}]_{ij} \quad \text{Eq. 14}$$

where $\mathbf{A}$ represents a matrix made by $A_{ij}$ introduced in Eq. 11. Figure 4 compares the calculated relative electric permittivity of PbTiO$_3$, calculated as $\varepsilon = (2\varepsilon_{11} + \varepsilon_{33})/3$, with the measured data reported by Remeika and Glass [32] and Bhide et al. [33].

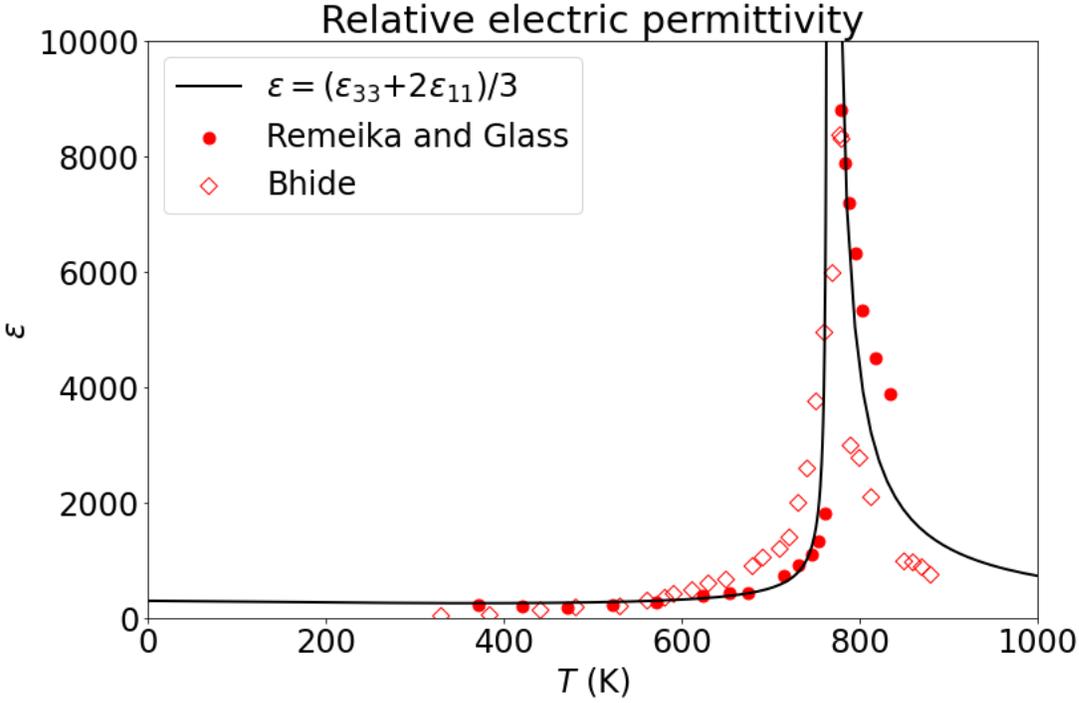

Figure 4. The relative electric permittivity of PbTiO$_3$. The lines represent the calculated results and the various symbols represent the measured data reported by Remeika and Glass [32] and Bhide et al. [33].

It is a common observation that phase transition can induce an extra heat capacity that mostly increases exponentially in wide range temperature range below the critical temperature. However, the extra heat capacity is often difficult to be predicted. We find that the extra heat capacity can be well described by the present approach for PbTiO$_3$. To facilitate the comparison with experimental data, the phonon contributions have been included by performing first-principles calculations ermploying the projector-augmented wave (PAW) method implemented in the Vienna *ab initio* simulation package (VASP, version 5.2) using the DFTTK package [34] with the LDA exchange-correlation functional [35]. To account for the effect of the dipole-dipole interaction on phonon frequency for insulators, the Born effective charge tensor and the high frequency static dielectric tensor needed in the mixed-space approach [36] are calculated by employing the linear-response theory as implemented in VASP 5.2 by Gajdos et al. [37].

Counting the contributions from lattice vibration and the extra Gibbs energy, the total heat capacity per atom is

$$C(T) = C_{vib}(T) + \frac{\Delta C(T)}{N_{PAC}}$$  Eq. 15

where $C_{vib}(T)$ is the lattice contribution to the heat capacity and the extra heat capacity $\Delta C(T) = d\Delta H(T)/dT$. The extra enthalpy is evaluated by $\Delta H(T) = \Delta G(T) + T\Delta S(T)$ and the extra entropy is evaluated by $\Delta S(T) = \partial \Delta G/\partial T$ where the temperature dependence of the order parameter can be accounted by Eq. 7. Here we note that $N_{PAC}$ in Eq. 15 is size the the PAC in the unit of number of atoms that can be determined through $N_{PAC} = 5V_{PAC}/V_{cell}$ where $V_{cell}$ (= ~62.4×10$^{-30}$ m³ from Haun et al. [28] at the the transition temperature) is the volume of perovskite unit cell of PbTiO$_3$. We find that $N_{PAC} = 10$ for PbTiO$_3$. Figure 5 shows the salient agreements between the calculations and experiments [27,38] for the heat capacities and the extra entropy between the calculations and experiments [27,38] as well as Landau fittings [27,38] in the temperature range below the transition. It is found that the Landau fittings become rather inaccurate when temperature is significant away (below ~400 K by Haun et a. [28] and below ~600 K by Rossetti and Maffei [27]) from the transition temperature. The present calculations support more the experiment by Yoshida et al. [38]. For example, the calculated extra entropy above the transition of ~1.46 J/mol-at/K exactly reproduced the reported value of Yoshida et al. [38].

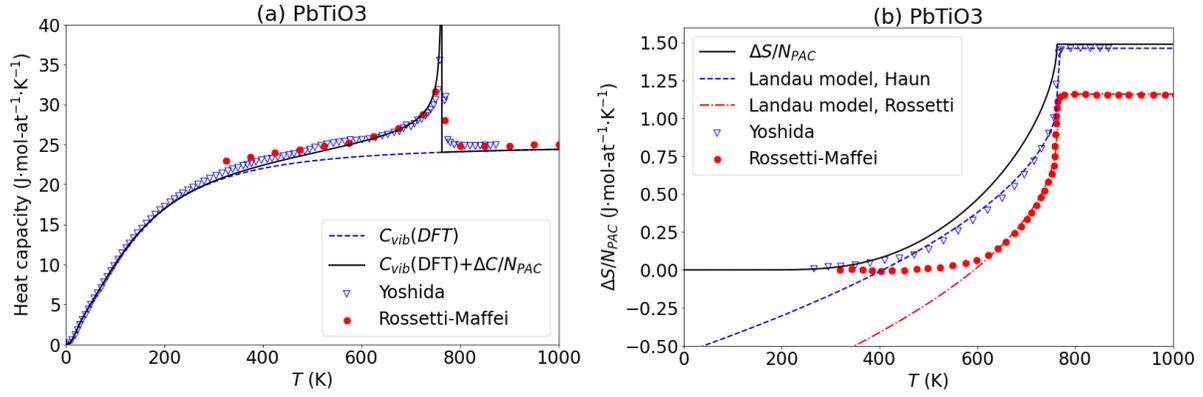

Figure 5. The calculated heat capacities (lines) and entropied (solid line) in comparison with experimental data (symbols) for PbTiO$_3$ [27,38]. (a) heat capacity; (b) extra entropy where the dashed and dot-dashed lines represent the Landau model fitting results by Haun et al. [28] and Rossetti and Maffei [27], respectively.

In summary, we demonstrate an approach for handling the general critical phenomena and phase transitions by Boltzmann thermal mixing among multiple parabolic potentials. We show that a macroscopic system with a complex free-energy landscape can be decomposed into an ensemble of multiple local parabolic wells. The approach is able to account for the lattice vibration and is applied to PbTiO$_3$ for the description of the ferroelectric-paraelectric transition and the associated thermodynamic and electric properties in the temperature range of 0-1000 K. The key theoretical contributions include: i) the extra free-energy is formulated using only the second order term of the order parameter, in comparison to a Landau polynomial expansion to the fourth or higher orders; ii) the long range field is derived as *a posteriori* result; iii) the approach naturally brings out the double-well and its temperature evolution to single-well without using *a priori* control of the temperature dependence of any parameter; iv) the extra free-energy is referenced to a single domain, uniform, and equilibrium ordered state whose free-energy can be well formulated and calculated within the framework of the first-principles theory, compared with that of Landau expansion which is referenced to high temperature structures and often causes issues of imaginary phonon modes in first-principles calculations.

**Acknowledgments**

This work was mainly supported by the Computational Materials Sciences Program funded by the US Department of Energy, Office of Science, Basic Energy Sciences, under Award Number DE-SC0020145 (Wang, Yang, and Chen); and partially supported by the National Science Foundation (NSF) with Grant No. CMMI-2050069 (Wang, Shang and Liu). First-principles calculations were performed partially on the Roar supercomputer at the Pennsylvania State University's Institute for Computational and Data Sciences (ICDS), partially on the resources of the National Energy Research Scientific Computing Center (NERSC) supported by the U.S. DOE Office of Science User Facility operated under Contract No. DE-AC02-05CH11231, and partially on the resources of the Extreme Science and Engineering Discovery Environment (XSEDE) supported by NSF with Grant No. ACI-1548562.